\documentclass{jpsj-suppl}
\usepackage{txfonts} %Please comment out this line unless the txfonts package is availabe in your LaTeX system.
\usepackage[hmargin=2.8cm,vmargin=3.0cm]{geometry}
\newcommand{\vect}[1]{\boldsymbol{#1}}

\title{A Search for non-Newtonian force in a precision measurement of the scattering of slow neutrons in Xenon gas}

\author{Yoshio \textsc{Kamiya}$^{1}$, Misato \textsc{Tani}$^{1}$, Sachio \textsc{Komamiya}$^{1}$, Guinyun \textsc{Kim}$^{2}$, and Kyungsuk \textsc{Kim}$^{2}$}

\inst{$^{1}$Department of Physics and International Center for Elementary Particle Physics, The University of Tokyo, Tokyo 113-0033, Japan \\
$^{2}$Department of Physics and Center for High Energy Physics, Kyungpook National University, Daegu 702-701, Korea}

\email{kamiya@icepp.s.u-tokyo.ac.jp}

\abst{
An experimental search for non-newtonian, gravity-like force in a precision measurement 
of the scattering of slow neutrons in Xenon gas is proposed.
A preliminary experiment with small statistics of 25 hours irradiation time was performed 
and the observed scattering distribution is consistent with the expectation with no additional forces.
A 95\% CL limit on the coupling strength for a hypothetical force of 1 nm interaction range was evaluated to be $2\times10^{-15}$.
The expected sensitivity for a planned high statistics runs is discussed.}

\kword{fifth force, non-Newtonian gravity, slow neutron}

\begin{document}
\maketitle

\section{Introduction}

\begin{figure}[htbp]
  \begin{center}
   \includegraphics[width=130mm]{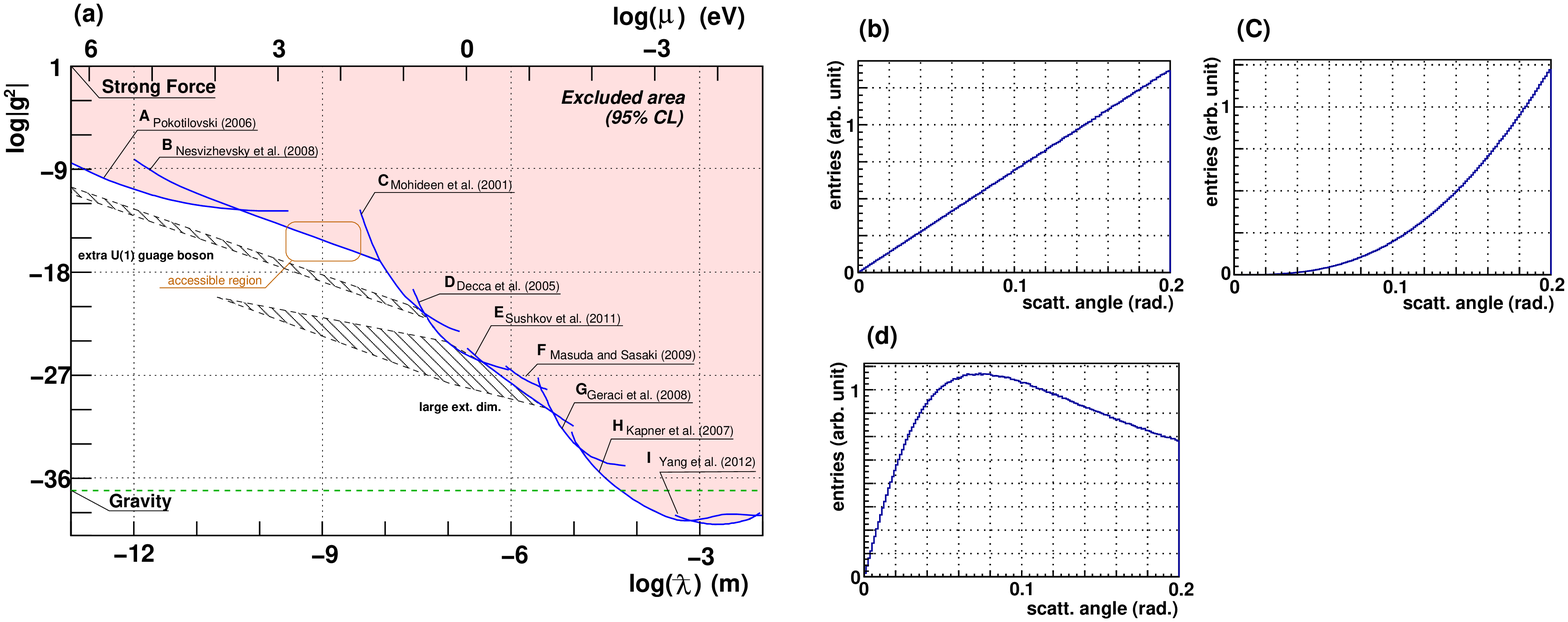}
  \end{center}
  \caption{(a) Current wold limits for non-Newtonian force
  in $g^2$-$\lambdabar$ space \cite{limitA, limitB, limitC, limitD, limitE, limitF, limitG, limitH, limitI}.
  The experimental scheme for the limits A and B are neutron scattering,
  and the limits C to I are obtained by a macroscopic scheme using a torsion balance or cantilever.
  Theoretical expectations from the large extra demotions\cite{ref.KK} and extra $U(1)$ gauge boson
  with supersymmetric extensions of the standard model\cite{ref.uboson} are shown as hatched areas.
  (b,c,d) Simulated scattering angle distribution of nuclear scattering, higher order EM scattering, 
  and additional Yukawa-type scattering ($\lambdabar = 1$ nm), respectively.
  The thermal motion of Xenon gas is considered in the simulation.
  }
  \label{fig.comb}
\end{figure}

``Is there any new force in addition to the four known fundamental forces''
``Is the Newtonian inverse square law of gravity valid even in the sub-mm range?''
These are still open questions, and some theories, such as the Kaluza-Klein theory with extra dimensions\cite{ref.KK},
predict the existence of a fifth force or show certain modifications to the Newtonian law of gravity.
A simple Yukawa-type parametrization of an additional scattering potential due to new physics is given by
\begin{equation}
V_{new}(r) = -\frac{1}{4\pi} g^2 Q_1 Q_2 \cdot \frac{e^{-\mu r}}{r} \hspace{1.5mm},
\end{equation}
where $g^2$ is a coupling strength, $Q_i$ are coupling charges, and $\mu$ is the mass of a mediating boson for the new force.
Much progress for the fifth force or the non-newtonian force search in this parametrization has been made over the last 10 years
and this field is still very active, especially for gravity-like interactions with coupling charge of mass number or macroscopic mass.
The current world limits in $g^2$-$\mu$ or $g^2$-$\lambdabar$ space are shown 
in figure \ref{fig.comb}(a)\cite{limitA, limitB, limitC, limitD, limitE, limitF, limitG, limitH, limitI}, where
$\lambdabar = 1/\mu$ is the interaction range.

This additional scattering potential slightly modifies the scattering angle distribution of the neutron-atom interaction.
We propose to search for new forces by evaluating a deviation of the angular distribution from the known scattering process.
The scheme has a sensitivity to new forces with around 1 nm interaction range, 
which corresponds to a mass of several hundred eV (see also Fig.\ref{fig.comb}(a)).
The current world limit at 1 nm range is about $7\times10^{-16}$, 
which is obtained by reviewing available neutron scattering data\cite{limitB}.
Statistical uncertainty limits the sensitivity.

%In this article, we show an experimental scheme to search for non-Newtonian force in a precision measurement of the scattering of 
%slow neutrons in Xenon gas. 
In this article, 
%we propose an experiment to search for non-Newtonian force in a precision measurement 
%of the scattering of slow neutrons in Xenon gas. 
%Our accessible region on the $g^2$-$\mu$ space is also shown in the figure \ref{fig.comb}(a).
we review the interaction between a slow neutron and Xenon
with an additional cross section due to a Yukawa-type potential.
After that, we show details of a preliminary experiment which was performed in January 2013 with small statistics of 25 hours irradiation time,
and the achievable sensitivity expected for a planned high statistics run.

\section{Review of slow neutron/Xenon interactions}
A neutron scattering length $b(\vect{q})$ for a diamagnetic atom such a Xenon
is derived from \cite{sears} as
\begin{equation}
b(\vect{q}) = b_c(q) + \frac{1}{\sqrt{I(I+1)}} \vect{\sigma}\cdot\vect{b_i}(\vect{q})\cdot\vect{I} + ib_s(q)\vect{\sigma}\cdot\vect{\hat{n}} \hspace{1.5mm} ,
\end{equation}
where $\vect{q}$ is the momentum transfer vector, $\vect{I}$ is the nucleus spin, 
$\vect{\sigma}/2$ is the neutron spin, and $\vect{\hat{n}}$ is a unit vector 
perpendicular to the plane of scattering. 
$b_c$ in the first term represents coherent scattering, the tensor $\vect{b_i}$ in the second term is for incoherent scattering,
and the last term of $b_s$ represents so-called Schwinger scattering, whose scattering plane is perpendicular to the one of the other scattering processes.
For an unpolarized neutron beam and Xenon atomic gas target, the second and the third terms vanish.
% as a function of the degree of polarizations.
The scales of each term, $b_c$, $\vect{b_i}$, and $b_s$ for the Xenon atom are $\sim 5$ fm, $\sim 10^{-3}$ fm, and $\sim 10^{-3}$ fm, respectively.
Therefore, even when the beam is polarized with a certain fraction,
the second and third terms can still be neglected in the following discussion for our intended sensitivity.

The coherent term is divided into two interactions, nuclear scattering and higher order electromagnetic scattering.
% that has no $q$ dependence 
%and higher order electromagnetic scattering whose angular distribution is empirically described by the atomic form factor $f(q)$.
The scattering length of the Yukawa-type potential is calculated by the Fourier transformation under the Born approximation and 
it is added to the known scattering lengths when a new force exists.
The differential cross section with the Yukawa-type scattering is described by the square of the total scattering length, and is written as
\begin{eqnarray}
\frac{d\sigma}{d\Omega} & \simeq & (b_{Nc} + b_{p})^2 \hspace{1mm} (1 + 2\chi[1-f(q)] + 2\chi_y[(\frac{q}{\mu})^2 + 1]^{-1} ) \\
f(q) & = & [1 + 3(\frac{q}{q_0})^{2}]^{-0.5} \hspace{5mm} {}_{(q_0 = 7 {\rm \AA}^{-1} \hspace{0.5mm} {\rm for \hspace{0.5mm} Xe})} \\
\chi & \equiv & \frac{b_F + b_I}{b_{Nc} + b_{p}}Z \hspace{5mm} \sim 10^{-2}\\
\chi_{y} & \equiv & \frac{m_n}{2\pi} g^2 Q_1 Q_2 \frac{1}{(b_{Nc} + b_{p})\mu^2}
\end{eqnarray}
where $b_{Nc}$ is the coherent nuclear scattering length, $b_P$ is the polarization scattering length, 
$b_F$ is the Foldy scattering length, $b_I$ is the intrinsic n-e scattering length, and $m_n$ is the neutron mass. 
$\chi$ and $\chi_y$ are small enough to omit their second order terms.
The first term from nuclear scattering has no $q$ dependence, on the other hand the second term
due to higher order electromagnetic scattering has $q$ dependence which is empirically described by the atomic form factor $f(q)$.
%The first, second, and third terms represent nuclear scattering, higher order electromagnetic scattering, and scattering by new force,
%called as nuclear term, n-e term, additional term, respectivly.
By integrating with azimuthal angle and by taking into account the thermal motion of Xenon gas,
the scattering angle $\theta$ distribution of each term is simulated 
as shown in figure \ref{fig.comb}(b), \ref{fig.comb}(c), \ref{fig.comb}(d) in the case of $\lambdabar=1$ nm.
These distributions are clearly distinguished each other, and,
thanks to this property, an analytical method using the shapes of the distributions can be adopted to avoid uncertainty on measuring absolute values 
such as the total coherent cross section. In later discussion, 
functions $h_1(\theta)$, $h_2(\theta)$, $h_3(\theta;\lambdabar)$ are used 
to express these simulated angular distributions shown in the figure \ref{fig.comb}(b), \ref{fig.comb}(c), \ref{fig.comb}(d), 
respectively.

\section{Preliminary Experiment}
\begin{figure}[htbp]
  \begin{center}
   \includegraphics[width=110mm]{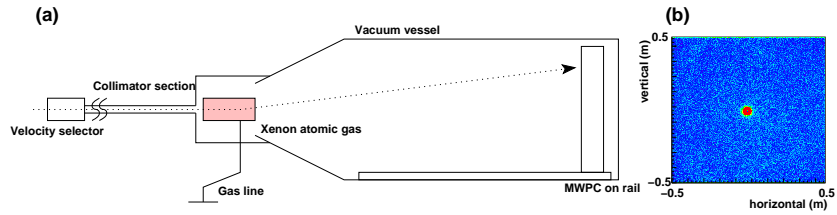}
  \end{center}
  \caption{(a) Schematic drawing of the preliminary experiment. 
  The total length from the velocity selector to the end of the vacuum vessel is 40 m.
  The detector can be moved in the vessel to adjust the distance from the sample.
  (b) Measured scattering image on the detector. The sensitive area is $ 1 \times 1$ m$^2$.}
    \label{fig.exp}
\end{figure}

A preliminary experiment was performed in Jan. 2013 with a 25 hour irradiation of a Xenon sample.
We use the 40 m SANS(Small Angle Neutron Scattering) beam line\cite{han} 
at the HANARO research reactor located at the Korean Atomic Energy Research Institute.
Figure \ref{fig.exp}(a) shows schematic drawing of the experiment.
The wavelength of neutrons was selected to be 5 $\AA$ with a 12\% FWHM spread.
The beam was collimated to have a 1 mrad divergence and the beam size was 12 mm in diameter.
The beam intensity was $1 \times 10^{5}$ neutrons/sec.
A chamber of 250 mm length was filled with Xenon gas at 1 atm pressure.
The detector used in this experiment was a $^{3}$He filled MWPC with 5 mm wire spacing along horizontal and vertical axes,
whose detection efficiency is around 80\% for neutrons with 5 $\AA$ wavelength.
The size of sensitive area is 1 m $\times$ 1 m square and the distance from the Xe chamber to the detector was 2.5 m.
The accuracy on measurements of the scattering angle is mainly limited by the finite chamber length,
and is better than 5\%. The other sources, due to the beam size and detector spatial resolution, give a negligible contribution.
The vacuum is required to be less than 1 Pa, considering the influence of scattering from residual gas.
A scattering image is shown in figure \ref{fig.exp}(b).

\begin{figure}[htbp]
  \begin{center}
   \includegraphics[width=115mm]{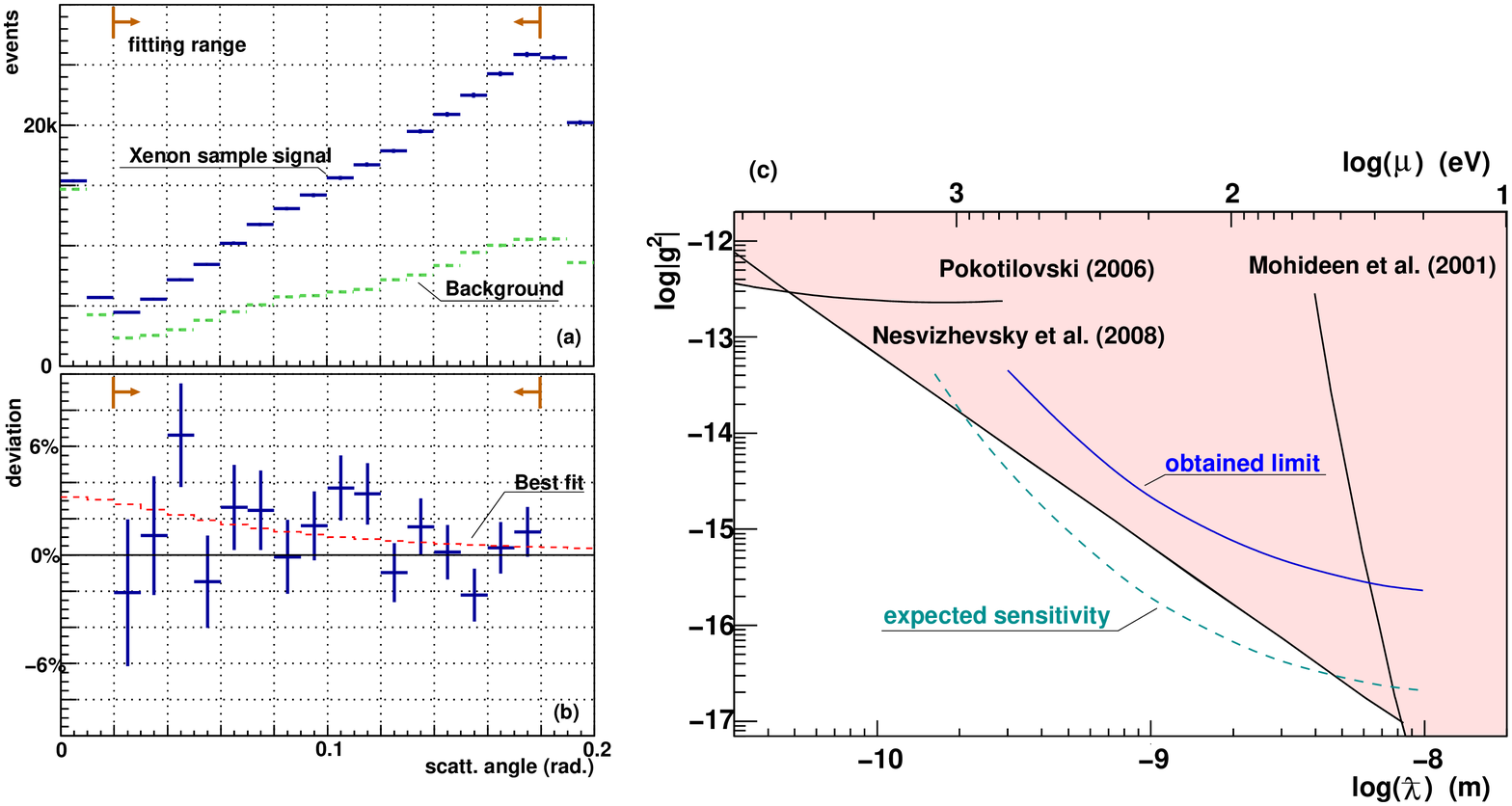}
  \end{center}
  \caption{(a) Scattering angle distribution for Xenon sample with background (solid line) and background only (dashed line).
  (b) Estimated deviation from the known scattering processes when the range of new force $\lambdabar = 1$ nm. 
  Data from 0.02 mrad to 1.8 mrad are used for fitting.
  (c) 95\% CL Limits obtained the preliminarily experiment (solid curve) and expected sensitivity for high statistic runs (dashed curve).}
  \label{fig.fit}
\end{figure}

In figure \ref{fig.fit}(a), two sets of scattering data with and without(background) Xenon gas are shown.
After subtracting the background distribution, the angular distribution of the Xenon sample signal 
$g(\theta)$ is evaluated by a least squares method with 
a linear combination of $h_1(\theta), h_2(\theta), h_3(\theta)$ written as
\begin{equation}
g(\theta) = (1-\alpha-\beta)h_1(\theta) + \alpha h_2(\theta) + \beta h_3(\theta; \lambdabar) \hspace{1mm},
\end{equation}
where $\alpha$ and $\beta$ are the parameters to be estimated, describing the fractions of each term.
The fitting range is from 0.02 to 0.18 mrad, to avoid the effects from the direct beam 
and the edge of the active area.
Figure \ref{fig.fit}(b) shows the analyzed deviations from the known scattering processes,
when the interaction range of a new force $\lambdabar$ is 1 nm.
The results is consistent with the expectation with no additional forces.

The 95\% CL limits for a Yukawa-type, non-Neutonian force obtained are $g^2 < 4.5\times10^{-15}$ 
for $\lambdabar = 0.3$ nm, $g^2 < 2\times10^{-15}$ for $\lambdabar = 1$ nm, and $g^2 < 2.7\times10^{-16}$ for 
$\lambdabar = 70$ nm. In figure \ref{fig.fit}(c), the limit is shown as a solid curve.
Each contribution to the systematic uncertainty $\delta_{g^2}$ in terms of $g^2$ estimation are evaluated using Monte Carlo pseudo-experiments:
for $\lambdabar = 1$ nm, 
$\delta_{g^2} \sim 5\times10^{-17}$ from a neutron flux uncertainty of 0.5\%;
$\delta_{g^2} \sim 2\times10^{-17}$ from detection non-uniformity of 1\%(RMS);
$\delta_{g^2} \sim 2\times10^{-17}$ from the neutron energy determination within 3\%;
$\delta_{g^2} < 10^{-17}$ from the atomic form factor model validity within 5\%;
$\delta_{g^2} < 10^{-17}$ from temperature uncertainty of $\pm$ 6 K; 
$\delta_{g^2} < 10^{-17}$ from the interaction position uncertainty of 25 cm due to the chamber size;
$\delta_{g^2} < 10^{-17}$ from the gas contamination of $10^{-5}$. 
The statistical uncertainty still dominats the experimental sensitivity of this scheme.
We are planning to perform high statistics runs with two order of magnitude larger statistics than this preliminary experiment.
The expected limits for this next experiment is shown as a dashed curve in figure \ref{fig.fit}(c).

%\subsection{Scattering Length}
%
%\subsubsection{Subsubsection}
%
%
%\begin{table}[tbh]
%\caption{Captions to tables and figures should be sentences.}
%\label{t1}
%\begin{tabular}{ll}
%\hline
%AAA & BBB \\
%CCC & DDD \\
%\hline
%\end{tabular}
%\end{table}
%
%\subsubsection{Equation numbers}
%
%The \verb|seceq| option resets the equation numbers at the start of each section.
%
%\begin{figure}[tbh]
%%\includegraphics{fig01.eps}
%\caption{You can embed figures using the \texttt{\textbackslash includegraphics} command. EPS is the only format that can be embedded. Basically, figures should appear where they are cited in the text. You do not need to separate figures from the main text when you use \LaTeX\ for preparing your manuscript.}
%\label{f1}
%\end{figure}
%
%Label figures, tables, and equations appropriately using the \verb|\label| command, and use the \verb|\ref| command to cite them in the text as ``\verb|as shown in Fig. \ref{f1}|". This automatically labels the numbers in numerical order.
%
%The \verb|minipage| environment can be used to place figures horizontally.
%
%\begin{equation}
%E = mc^{2}
%\label{e1}
%\end{equation}
%
%\appendix
%\section{}
%
%Use the \verb|\appendix| command if you need an appendix(es). The \verb|\section| command should follow even though there is no title for the appendix (see above in the source of this file).

\section*{Acknowledgment}
The authors are grateful to Dr. Tae-Hwan Kim and Dr. Young-Soo Han for their specialist support in maintaining beam conditions,
and Dr. Jong-Dae Jang and Dr. Eun-Hye Kim for supporting beam operation. 
This work is supported by JSPS KAKENHI Grant No. 90434323.

\end{document}